\documentclass[sigconf,nonacm]{acmart}

\acmConference[EASE 2026]{The 30th International Conference on Evaluation and Assessment in Software Engineering}{9--12 June, 2026}{Glasgow, Scotland, United Kingdom}

\usepackage{graphicx}
\usepackage{booktabs}
\usepackage{array}
\usepackage{xcolor}
\usepackage{colortbl}

\newif\ifshowchanges
\showchangesfalse

\newcommand{\revised}[1]{%
  \ifshowchanges{\color{blue}#1}\else{#1}\fi%
}
\usepackage{makecell}
\usepackage{microtype}
\usepackage{balance}
\usepackage{algorithm}
\usepackage{algpseudocode}
\usepackage{amsmath}
\usepackage{longtable}
\usepackage{enumitem}
\usepackage{tcolorbox}
\tcbuselibrary{breakable, skins}

\graphicspath{{figures/}}

\usepackage{listings}
\usepackage{listingsutf8}
\usepackage{listings}
\usepackage{listingsutf8}

\lstdefinestyle{promptstyle}{
  basicstyle=\scriptsize\ttfamily,
  breaklines=true,
  breakatwhitespace=true,
  columns=fullflexible,
  keepspaces=true,
  frame=single,
  framerule=0.5pt,
  rulecolor=\color{hdrblue},
  backgroundcolor=\color{rowp01},
  xleftmargin=4pt,
  xrightmargin=4pt,
  aboveskip=4pt,
  belowskip=4pt,
  showstringspaces=false,
  upquote=true
}

\definecolor{rowp01}{RGB}{234,242,251}
\definecolor{rowp02}{RGB}{254,249,240}
\definecolor{rowodd}{RGB}{240,244,248}
\definecolor{rowgreen}{RGB}{226,239,218}
\definecolor{promptbg}{RGB}{248,250,252}
\definecolor{promptborder}{RGB}{100,130,180}
\definecolor{promptheader}{RGB}{46,64,87}
\definecolor{hdrblue}{RGB}{46,64,87}

\tcbset{
  promptbox/.style={
    breakable, enhanced,
    colback=promptbg, colframe=promptborder,
    fonttitle=\bfseries\small\color{white}, coltitle=white,
    attach boxed title to top left={yshift=-2mm, xshift=4mm},
    boxed title style={colback=promptheader, colframe=promptheader, rounded corners, size=small},
    arc=3pt, boxrule=0.8pt,
    left=4pt, right=4pt, top=8pt, bottom=4pt,
    before skip=6pt, after skip=6pt,
    fontupper=\small,
  }
}

\begin{document}

\title{Prompt Engineering Strategies for LLM-based Qualitative Coding\\
of Psychological Safety in Software Engineering Communities:\\
A Controlled Empirical Study}

\author{Moaath Alshaikh}
\orcid{0000-0002-6484-4486}
\affiliation{%
  \institution{Federal University of Bahia}
  \city{Salvador}
  \state{Bahia}
  \country{Brazil}}
\email{moaathalshaikh@ufba.br}

\author{Tasneem Alshaher}
\orcid{0009-0000-1336-1671}
\affiliation{%
  \institution{Federal University of Bahia}
  \city{Salvador}
  \state{Bahia}
  \country{Brazil}}
\email{tasneemalshaher@ufba.br}

\author{Ricardo Vieira}
\orcid{0009-0007-1927-7957}
\affiliation{%
  \institution{Federal University of Bahia}
  \city{Salvador}
  \state{Bahia}
  \country{Brazil}}
\email{revieira@ufba.br}

\author{Beatriz Santana}
\orcid{0000-0003-4048-1921}
\affiliation{%
  \institution{Federal University of Bahia}
  \city{Salvador}
  \state{Bahia}
  \country{Brazil}}
\email{santanab@ufba.br}

\author{Clelio Xavier}
\orcid{0009-0008-5199-8487}
\affiliation{%
  \institution{Federal University of Bahia}
  \city{Salvador}
  \state{Bahia}
  \country{Brazil}}
  \email{clelio.xavier@ufba.br}

\author{Jose Amancio}
\orcid{0000-0002-9509-5238}
\affiliation{%
  \institution{State University of Feira de Santana}
  \city{Feira de Santana}
  \state{Bahia}
  \country{Brazil}}
\email{zeamancio@uefs.br}

\author{Glauco Carneiro}
\orcid{0000-0001-6241-1612}
\affiliation{%
  \institution{Federal University of Sergipe}
  \city{Aracaju}
  \state{Sergipe}
  \country{Brazil}}
\email{glauco.carneiro@dcomp.ufs.br}

\author{Julio Leite}
\orcid{0000-0002-0355-0265}
\affiliation{%
  \institution{Federal University of Bahia}
 \city{Salvador}
  \state{Bahia}
  \country{Brazil}}
\email{julioleite.ufba@gmail.com}

\author{Savio Freire}
\orcid{0000-0002-3989-9442}
\affiliation{%
 \institution{Federal Institute of Ceara}
 \city{Morada Nova}
 \state{Cear\'a}
 \country{Brazil}}
\email{savio.freire@ifce.edu.br}

\author{Manoel Mendonca}
\orcid{0000-0002-0874-7665}
\affiliation{%
  \institution{Federal University of Bahia}
  \city{Salvador}
  \state{Bahia}
  \country{Brazil}}
\email{manoel.mendonca@ufba.br}

\renewcommand{\shortauthors}{Alshaikh et al.}

\begin{abstract}
Qualitative analysis plays a pivotal role in understanding the human and social aspects of software engineering. However, it remains a demanding process shaped by the subjective interpretation of individual researchers and sensitive to methodological choices such as prompt design. Recent advancements in Large Language Models (LLMs) offer promising opportunities to support this type of analysis, although their reliability in reproducing human qualitative reasoning under varying prompting conditions remains largely untested. This study presents a controlled empirical evaluation of three LLMs---Claude Haiku, DeepSeek-Chat, and Gemini 2.5 Flash---across two prompt engineering strategies (zero-shot and multi-shot closed coding), using Cohen's $\kappa$ as the primary agreement metric over ten independent runs per configuration. Results suggest that multi-shot prompting significantly improves agreement for Claude Haiku ($\Delta\kappa = +0.034$, Wilcoxon $p = 0.004$) but not for DeepSeek-Chat or Gemini 2.5 Flash. Intra-model stability varies substantially---DeepSeek-Chat and Claude Haiku exhibit the lowest variance ($\text{SD} \approx 0.017$), while Gemini 2.5 Flash is the least stable ($\text{SD} = 0.038$). A systematic over-prediction of ``Sharing Negative Feedback'' is identified across all models (bias ratios up to $5.25\times$), alongside consistent under-prediction of ``Expressing Concerns.'' Collectively, these findings provide empirical evidence for prompt engineering guidelines in LLM-assisted qualitative coding for software engineering research.
\end{abstract}

\ccsdesc[300]{Software and its engineering~Empirical software validation}
\ccsdesc[500]{Computing methodologies~Natural language processing}

\keywords{Prompt Engineering, LLMs, Qualitative Coding, Psychological Safety, Software Engineering}

\maketitle

{\footnotesize
\noindent\textit{
Preprint version of a paper accepted at the
\href{https://conf.researchr.org/home/ease-2026/prompt-se-2026\#event-overview}
{1st International Workshop on Prompt Engineering for Software Engineering (PROMPT-SE 2026)},
co-located with the 30th International Conference on Evaluation and Assessment in Software Engineering (EASE 2026),
Glasgow, Scotland, United Kingdom, June 9--12, 2026.
}
}
\section{Introduction}

Large language models (LLMs) are among the most prominent recent developments in artificial intelligence, emerging as promising tools for qualitative text analysis and interpretation across academic disciplines~\cite{barros2025}. In software engineering (SE), the growing interest in understanding human factors has intensified the adoption of qualitative methods to explore social phenomena such as psychological safety, team communication, and interpersonal conflict~\cite{seaman2025}. LLMs have been applied in this domain to analyze professional programming forums and support evidence-based reasoning~\cite{leca2025}, yet their reliability under varying prompting conditions remains largely untested.

Prompt engineering has emerged as a vital research area for ensuring reliable results from language models~\cite{liu2023}, yet current empirical evidence on how specific prompting strategies affect qualitative coding performance is limited. Questions of direct practical relevance---whether zero-shot prompting is sufficient, whether multi-shot examples improve agreement, whether models exhibit systematic category bias, and whether LLM outputs are stable across repeated runs---remain unanswered in the context of qualitative coding for SE.

This study addresses this gap through a controlled empirical evaluation of three state-of-the-art LLMs---Claude Haiku, DeepSeek-Chat, and Gemini 2.5 Flash---across two prompt configurations applied to a human-coded Gold Standard of 116 quotes related to psychological safety in SE communities on Stack Exchange. Each configuration was executed ten independent times per model to measure both mean agreement and intra-model stability---a dimension frequently overlooked in comparable studies. Four research questions guide this study:

\begin{itemize}[leftmargin=*,noitemsep]
  \item \textbf{RQ1:} To what extent can LLMs reproduce human qualitative coding of psychological safety statements in software engineering discussions, as measured by Cohen's $\kappa$ and per-class F1 scores?
  \item \textbf{RQ2:} How does prompt design (zero-shot vs.\ multi-shot) influence the reliability of LLM-based qualitative coding?
  \item \textbf{RQ3:} How stable are LLM qualitative coding results across repeated runs under the same prompt configuration?
  \item \textbf{RQ4:} Do LLMs exhibit systematic bias when coding psychological safety categories compared with human analysts?
\end{itemize}

\section{Related Work}

Barros et al.~\cite{barros2025} conducted a systematic mapping study demonstrating that LLMs can substantially improve both efficiency and analytical accuracy in qualitative research. Within SE, Bano et al.~\cite{bano2024} found persistent concerns around bias, transparency, and interpretive reliability in LLM-assisted qualitative analysis. Lenberg et al.~\cite{lenberg2024} provided methodological guidelines emphasizing the balance between automation and human interpretation. Gao et al.~\cite{gao2024} introduced CollabCoder, demonstrating how LLMs can strengthen analytical consistency in cooperative qualitative analysis. Le\c{c}a et al.~\cite{leca2025} described LLMs as a ``new frontier'' for empirical SE.

On the prompt engineering side, Liu et al.~\cite{liu2023} provided a systematic survey of prompting methods including zero-shot and multi-shot configurations. Wei et al.~\cite{wei2022} demonstrated that chain-of-thought prompting can elicit reasoning in LLMs. White et al.~\cite{white2023} proposed a catalog of prompt patterns for structured classification tasks. Chen et al.~\cite{chen2025} examined prompt engineering in terms of fairness and reproducibility. Despite these advances, most existing studies evaluate LLM performance in a single run without assessing cross-run stability, and few compare multiple models under identical conditions on the same human-coded dataset---gaps this study directly addresses.

\section{Research Method}

\subsection{Reference Dataset}

The analysis relies on a human-coded reference dataset compiled by Santana et al.~\cite{santana2023}, including 116 quotes from the Stack Exchange Project Management (SEPM) and Software Engineering (SESE) communities. All quotes were manually classified into seven behavioral codes from Edmondson's framework~\cite{edmondson1999}: \textit{Admitting Mistakes} (AM, $n=3$), \textit{Disagreeing with Suggestions/Ideas} (DS, $n=35$), \textit{Drawing Attention to Errors} (DAE, $n=5$), \textit{Expressing Concerns} (EC, $n=50$), \textit{Recommending Changes} (RC, $n=14$), \textit{Seeking Help} (SH, $n=5$), and \textit{Sharing Negative Feedback} (SNF, $n=4$). The label distribution is markedly imbalanced---EC accounts for 43.1\% of all quotes---motivating the bias analysis in Section~\ref{sec:results}. \revised{Santana et al.\ employed a multi-researcher coding process with consensus resolution of divergences; the authors report a 4.67\% inter-researcher disagreement rate, though no formal inter-rater reliability statistic (e.g., Cohen's~$\kappa$) is provided in the original study.}

\subsection{Methodological Workflow}

\begin{figure}[t]
  \centering
  \includegraphics[width=\columnwidth]{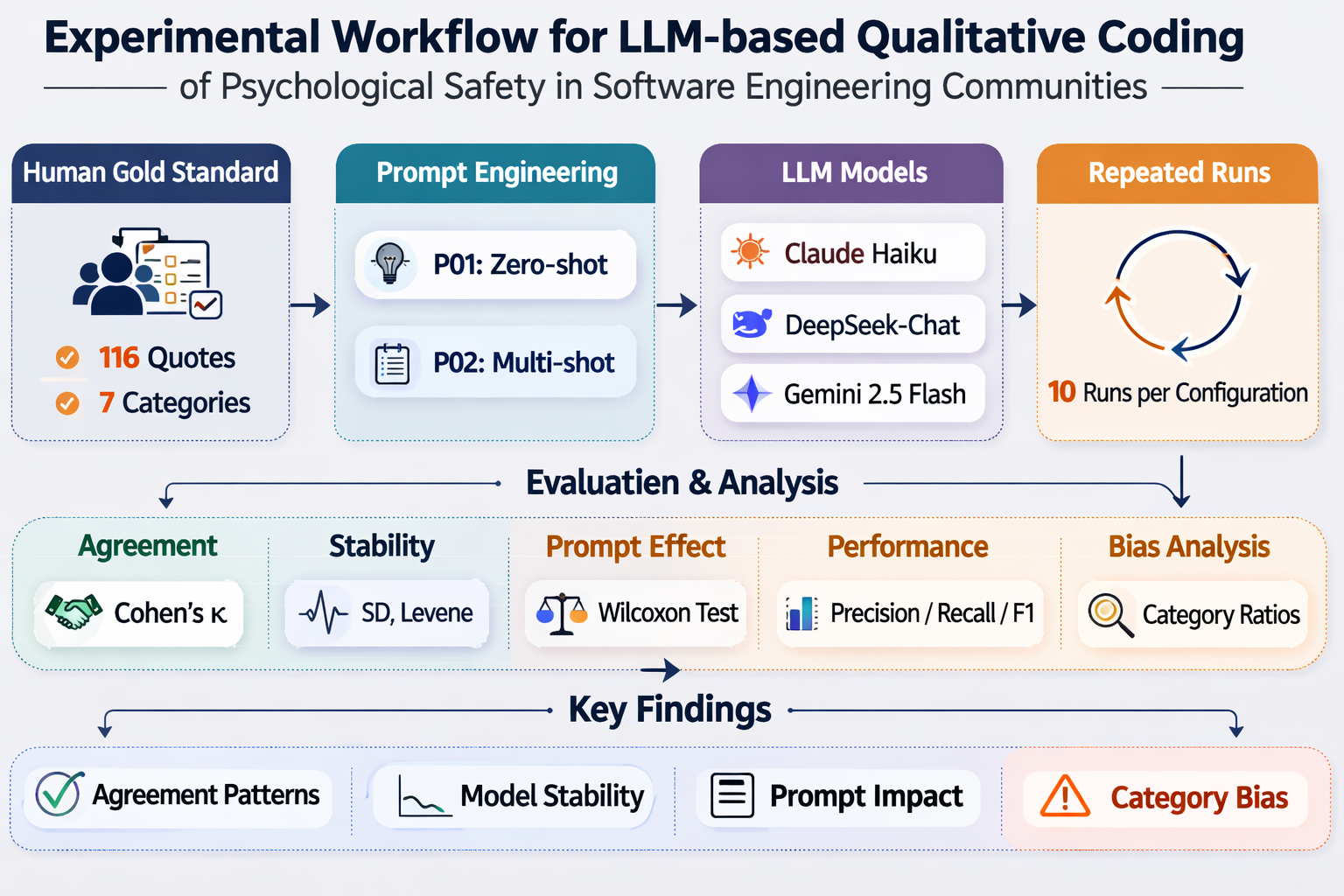}
  \caption{LLM-based Exploratory Workflow for LLM-based Qualitative Coding of Psychological Safety in Software \revised{Engineering} Communities}
  \label{fig:workflow}
\end{figure}

Figure~\ref{fig:workflow} illustrates the overall methodological workflow, which proceeds through four stages: (1) dataset preparation and gold standard establishment, (2) prompt engineering design, (3) automated LLM coding with repeated runs, and (4) statistical evaluation. Algorithm~\ref{alg:workflow} provides procedural pseudocode for each configuration across all three models and ten independent runs.

\begin{algorithm}[t]
\caption{LLM-based Exploratory Workflow with Human Oversight}
\label{alg:workflow}
\footnotesize
\begin{algorithmic}[1]
\Require 116 quotes; 7 codes~\cite{edmondson1999}; Models = \{Claude Haiku, DeepSeek-Chat, Gemini 2.5 Flash\}; Prompts = \{P01: zero-shot, P02: multi-shot\}; GoldStandard~\cite{santana2023}; Runs = 10
\Ensure $\kappa$ scores, stability metrics, bias indices
\State \textbf{P01 -- Zero-Shot:} classify 116 quotes $\times$ 3 models $\times$ 10 runs (0 example/label)
\State Compute $\kappa(\text{LLM}_r, \text{GS})$; record per-class F1 scores; compute category bias ratios
\State \textbf{P02 -- Multi-Shot:} repeat with 7 examples/label $\times$ 3 models $\times$ 10 runs
\State Wilcoxon($\kappa_{\text{P01}}, \kappa_{\text{P02}}$); Levene variance test; cross-model comparison
\end{algorithmic}
\end{algorithm}

\section{Experimental Design and Setup}

\subsection{Prompt Engineering Design}

Prompt engineering served as both an analytical control and a bias-mitigation mechanism~\cite{bang2023,liu2023,schulhoff2024prompt}. Each prompt follows a standardized schema: role assignment, psychological safety definition, step-by-step instructions, code definitions, illustrative examples, and a structured output format (ID--Category--Challenge--Quote). The key design differences are: \textbf{P01} provides 0 example per code (zero-shot baseline); \textbf{P02} provides 7 examples per code (multi-shot enrichment). Complete prompt texts are {provided in Appendix~\ref{app:prompt}. Table~\ref{tab:prompts} summarizes the two configurations evaluated. \revised{P01 provides no annotated examples---only category names and decision rules---and therefore constitutes a true zero-shot configuration. Any earlier reference to ``1~example/code'' described a definitional clause within the category description, not an annotated training example.}

\begin{table}[t]
\centering
\caption{Prompt Configurations Used in Closed-Coding Phases}
\label{tab:prompts}
\footnotesize
\setlength{\tabcolsep}{3pt}
\begin{tabular}{@{}p{0.55cm}p{1.7cm}p{2.8cm}p{2.2cm}@{}}
\toprule
\textbf{ID} & \textbf{Phase} & \textbf{Key Design Choice} & \textbf{Scope} \\
\midrule
\rowcolor{rowp01}
P01 & Zero-Shot & \revised{No annotated examples; category names \& classification rules only}; 7 labels available & All 3 models; 116 quotes \\
P02 & Multi-Shot & 7 examples/code; 7 labels available & All 3 models; 116 quotes \\
\bottomrule
\end{tabular}
\end{table}

\subsection{Experimental Setup}
We evaluated: Claude Haiku (\texttt{claude-haiku-4-5-20251001}), \revised{Deep-Seek-Chat (\texttt{deepseek-chat}, accessed via \texttt{api.deepseek.com}, Open-AI-compatible endpoint),} and Gemini 2.5 Flash (\texttt{gemini-2.5-flash}). Each was evaluated on both P01 and P02 over ten independent runs, yielding 60 classification runs in total ($3 \times 2 \times 10$). Repeated runs are necessary because LLM outputs may vary across executions due to stochastic sampling processes. \revised{All three models were queried with \texttt{temperature~=~0} to minimise stochastic variation; no \texttt{seed} parameter was applied (the parameter is unsupported by the Gemini~API and was not set for Claude~Haiku or DeepSeek-Chat). No \texttt{top\_p} or \texttt{frequency\_penalty} overrides were applied beyond API defaults.}


\subsection{Evaluation Metrics}

\textbf{Cohen's $\kappa$ (Agreement):} Per-run agreement corrected for chance; aggregated as mean $\pm$ SD across ten runs. Interpreted using the Landis--Koch scale: $< 0.20$ Slight, 0.20--0.40 Fair, 0.40--0.60 Moderate, 0.60--0.80 Substantial.

\textbf{Intra-Model Stability (SD of $\kappa$):} Variance of $\kappa$ across ten independent runs. Levene's test compares variance between models at each configuration.

\textbf{Bias Ratio:} Ratio of LLM-predicted count for a given category to its human-coded Gold Standard count. Ratio $= 1.00$ indicates no bias; values $> 1.20$ flag over-prediction; values $< 0.80$ flag under-prediction. \revised{Note that the Bias Ratio captures distributional volume---the aggregate predicted count versus the gold-standard count---and does not reflect per-instance classification accuracy. It should therefore be interpreted alongside per-class F1 scores rather than in isolation.}

Per-class F1 scores (majority vote, ten runs) and Wilcoxon signed-rank test (P01 vs.\ P02, $n=10$ pairs, with Cohen's $d$ as effect size) complement the primary metrics.

\section{Results}
\label{sec:results}

\subsection{RQ1 -- Agreement Between LLM Coding and Human Coding}

Figure~\ref{fig:kappa} shows the $\kappa$ distribution across ten runs for all models under both prompt configurations. Table~\ref{tab:kappa} summarizes mean $\kappa$, SD, and range.

\begin{figure}[t]
  \centering
  \includegraphics[width=\columnwidth]{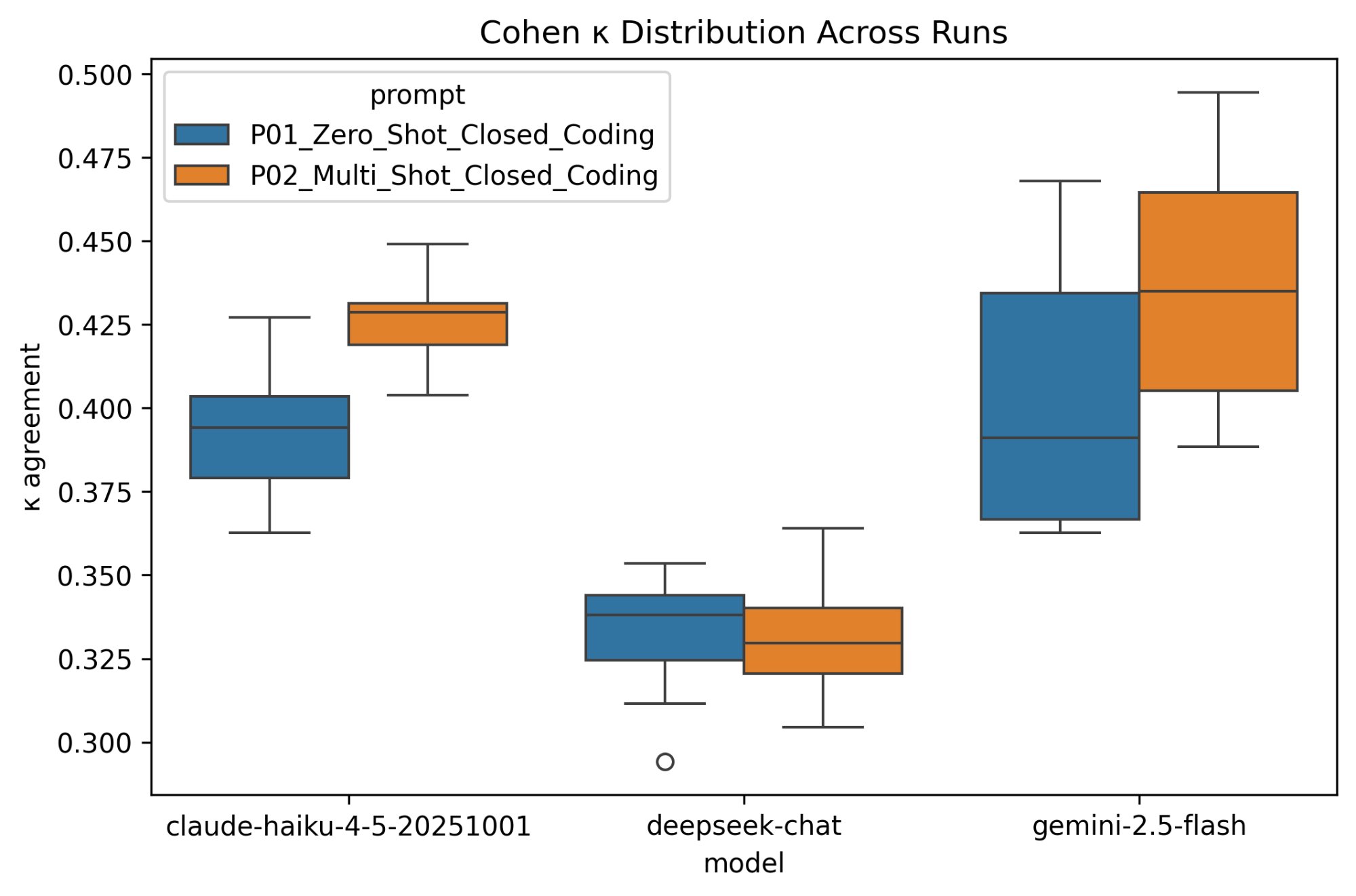}
  \caption{Cohen's $\kappa$ distribution across ten independent runs for all three models under P01 (zero-shot) and P02 (multi-shot). Boxes show interquartile range; whiskers extend to 1.5$\times$IQR; circles denote outliers.}
  \label{fig:kappa}
\end{figure}

\begin{table}[t]
\centering
\caption{Cohen's $\kappa$ Results --- All Configurations ($N=116$, 10 runs)}
\label{tab:kappa}
\footnotesize
\setlength{\tabcolsep}{3pt}
\begin{tabular}{@{}llccccc@{}}
\toprule
\textbf{Prompt} & \textbf{Model} & $\bar{\kappa}$ & \textbf{SD} & \textbf{Min} & \textbf{Max} & \textbf{Interp.} \\
\midrule
\rowcolor{rowp01}
P01 & Claude Haiku     & 0.392 & 0.018 & 0.363 & 0.427 & Fair \\
\rowcolor{rowp01}
P01 & DeepSeek-Chat    & 0.332 & 0.017 & 0.294 & 0.354 & Fair \\
\rowcolor{rowp01}
P01 & Gemini 2.5 Flash & 0.403 & 0.038 & 0.363 & 0.468 & Fair \\
\midrule
\rowcolor{rowp02}
P02 & Claude Haiku     & \textbf{0.426} & 0.011 & 0.404 & 0.449 & Moderate \\
\rowcolor{rowp02}
P02 & DeepSeek-Chat    & 0.331 & 0.017 & 0.304 & 0.364 & Fair \\
\rowcolor{rowp02}
P02 & Gemini 2.5 Flash & 0.437 & 0.035 & 0.388 & 0.495 & Moderate \\
\bottomrule
\end{tabular}
\end{table}

Overall, the results suggest fair to moderate agreement between LLM-generated labels and the human gold standard. Under P01, $\kappa$ values ranged from 0.332 (DeepSeek-Chat) to 0.403 (Gemini 2.5 Flash), all within the Fair range. Under P02, Claude Haiku and Gemini 2.5 Flash crossed into the Moderate range ($\kappa = 0.426$ and $0.437$ respectively), while DeepSeek-Chat remained essentially unchanged ($\kappa = 0.331$).

Per-class F1 scores are illustrated in Figure~\ref{fig:f1}. Across all models, ``Disagreeing with Suggestions or Ideas'' achieves the highest F1 scores (0.58--0.70), while minority categories such as ``Sharing Negative Feedback'' ($\text{F1} = 0.21$--$0.30$) and ``Admitting Mistakes'' ($\text{F1} \approx 0.33$--$0.40$) are systematically under-predicted. These patterns confirm that LLMs can partially reproduce human coding of common categories but struggle with semantically rare or overlapping ones. \revised{Caution is warranted when interpreting F1 scores for minority categories (AM: $n=3$; SNF: $n=4$; DAE: $n=5$): with support sizes this small, a single misclassification shifts F1 by 0.10--0.33 points, rendering these values indicative rather than statistically reliable.}

\begin{figure}[t]
  \centering
  \includegraphics[width=\columnwidth]{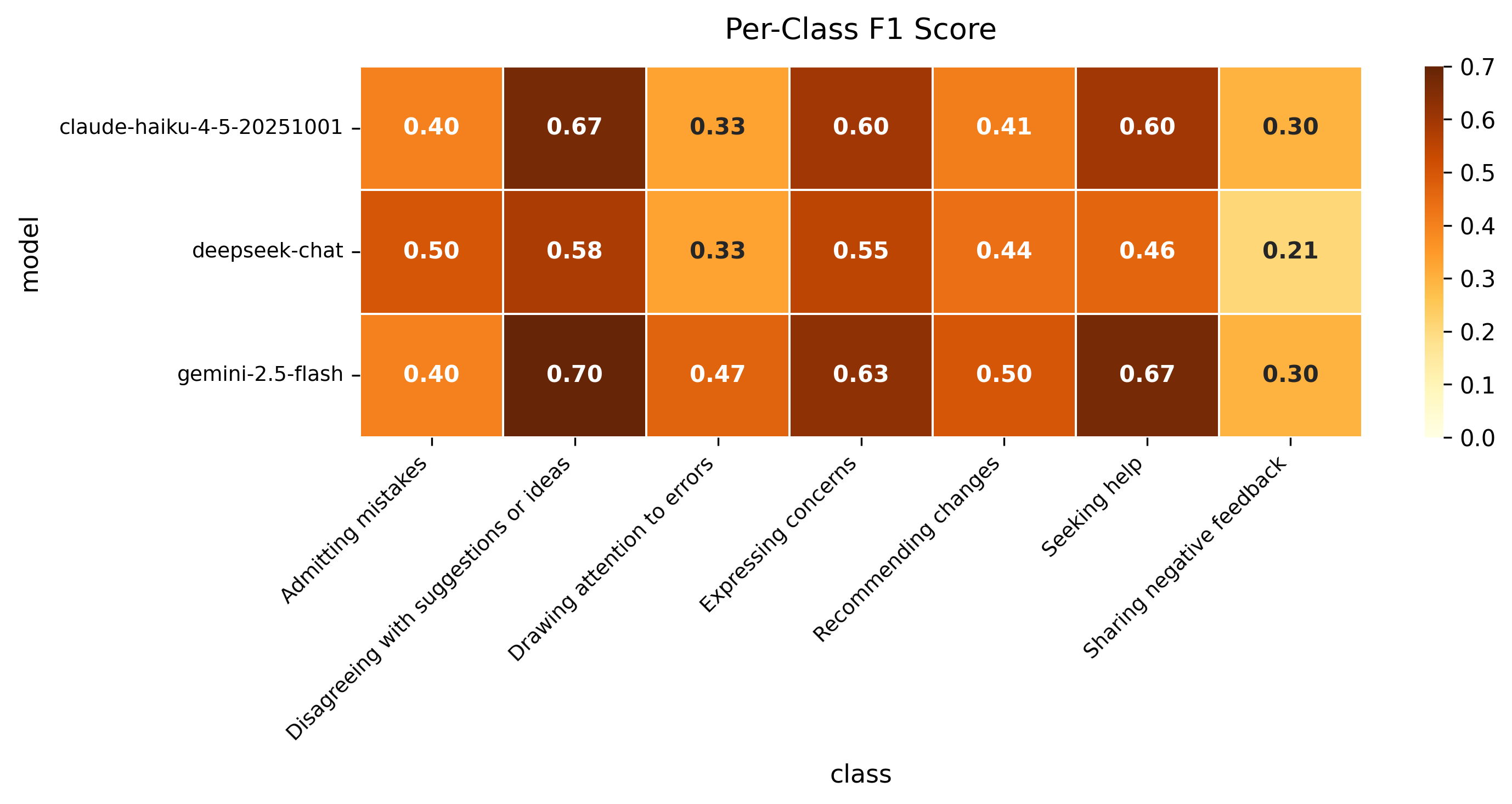}
  \caption{Per-class F1 scores (majority vote, 10 runs) aggregated across both prompt configurations. Color scale: green = high F1, red = low. ``Disagreeing with Suggestions or Ideas'' achieves the highest F1; ``Sharing Negative Feedback'' is consistently under-predicted. \revised{F1 scores for minority categories (AM: $n=3$; SNF: $n=4$; DAE: $n=5$) should be treated as indicative only due to very small support sizes.}}
  \label{fig:f1}
\end{figure}

\subsection{RQ2 -- Effect of Prompt Engineering}

Table~\ref{tab:wilcoxon} reports Wilcoxon signed-rank test results comparing $\kappa$ distributions between P01 and P02 for each model.

\begin{table}[t]
\centering
\caption{Wilcoxon Signed-Rank Test --- P01 vs.\ P02 per Model (RQ2)}
\label{tab:wilcoxon}
\footnotesize
\setlength{\tabcolsep}{3.5pt}
\begin{tabular}{@{}lrrrrrr@{}}
\toprule
\textbf{Model} & $\bar{\kappa}_{\text{P01}}$ & $\bar{\kappa}_{\text{P02}}$ & $\Delta\kappa$ & W & $p$ & $d$ \\
\midrule
Claude Haiku     & 0.392 & 0.426 & $+0.034$ & 1.0  & \textbf{0.004} & 2.41 \\
DeepSeek-Chat    & 0.332 & 0.331 & $-0.001$ & 25.0 & 0.846          & $-0.02$ \\
Gemini 2.5 Flash & 0.403 & 0.437 & $+0.033$ & 10.0 & 0.084          & 0.89 \\
\bottomrule
\multicolumn{7}{l}{\textit{$n=10$ pairs per test.}}
\end{tabular}
\end{table}

Claude Haiku showed a statistically significant improvement from P01 to P02 ($\Delta\kappa = +0.034$, $p = 0.004$, $d = 2.41$), crossing the Landis--Koch boundary from Fair to Moderate. \revised{The large Cohen's $d$ (2.41) reflects near-perfect directional consistency across all ten run-pairs rather than the magnitude of the mean difference; $d$ computed on small paired samples ($n=10$) is sensitive to low within-pair variance, and when all ten P02 runs consistently exceed their P01 counterparts by a small but stable margin, $d$ inflates relative to the mean difference---a known behaviour of effect-size estimates on small paired samples.} This large effect size suggests that multi-shot prompting meaningfully benefits models that are already in the Fair range. Gemini 2.5 Flash showed a similar positive trend ($\Delta\kappa = +0.033$, $p = 0.084$), marginally missing the $p < 0.05$ threshold. DeepSeek-Chat showed virtually no change ($\Delta\kappa = -0.001$, $p = 0.846$), suggesting its coding behavior is largely insensitive to example count in this closed-coding task.

\subsection{RQ3 -- Stability of LLM Coding Across Runs}

Figure~\ref{fig:stability} and Table~\ref{tab:stability} report the standard deviation of $\kappa$ across ten runs per configuration.

\begin{figure}[t]
  \centering
  \includegraphics[width=\columnwidth]{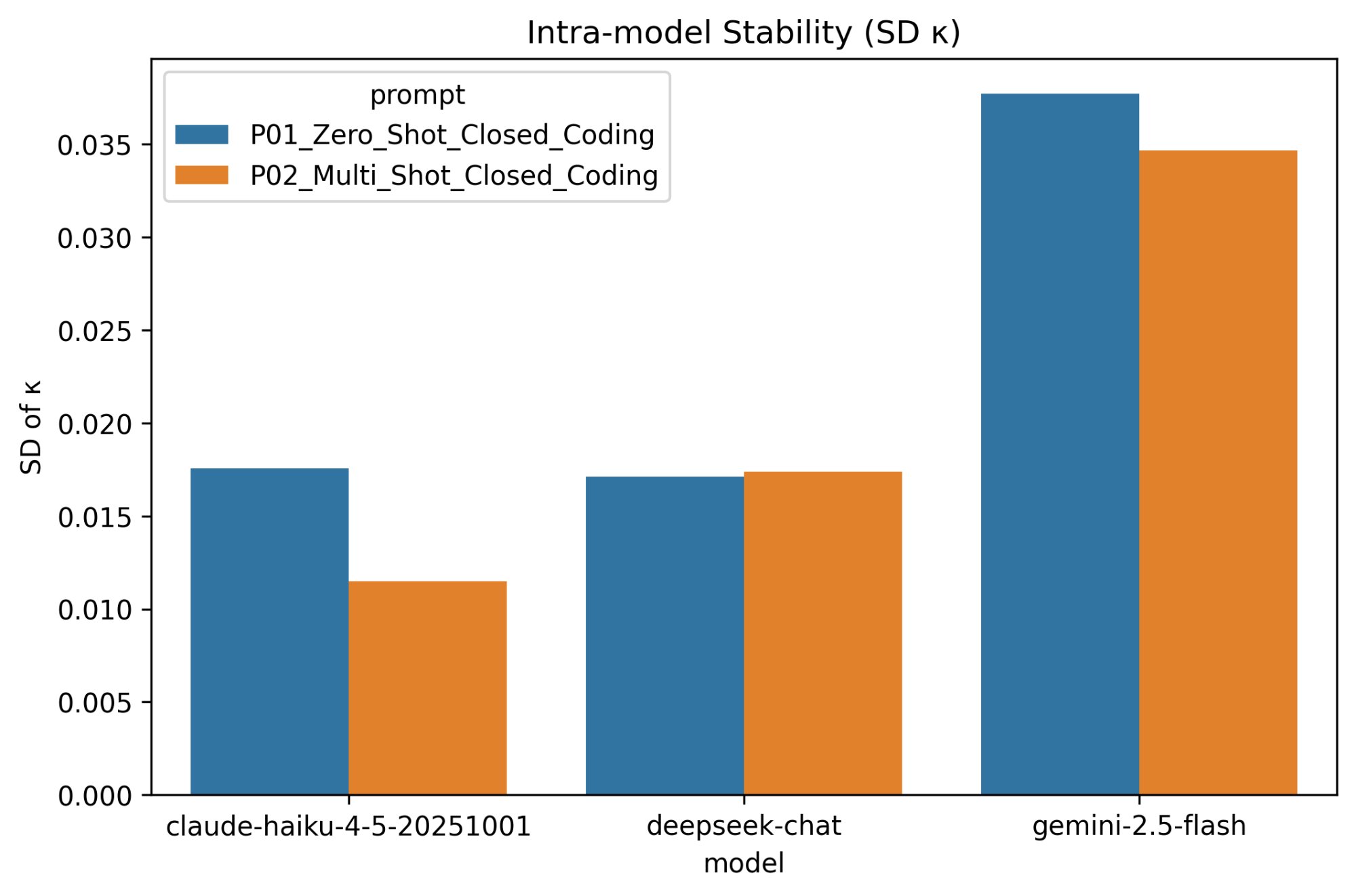}
  \caption{Intra-model stability: SD of $\kappa$ across 10 runs per configuration. Lower values indicate more reproducible coding behavior. Multi-shot prompting reduces variance for Claude Haiku.}
  \label{fig:stability}
\end{figure}

\begin{table}[t]
\centering
\caption{Intra-Model Stability --- SD of $\kappa$ Across 10 Runs (RQ3)}
\label{tab:stability}
\footnotesize
\setlength{\tabcolsep}{4pt}
\begin{tabular}{@{}lccc@{}}
\toprule
\textbf{Model} & SD$_{\text{P01}}$ & SD$_{\text{P02}}$ & \textbf{Direction} \\
\midrule
Claude Haiku      & 0.018 & \textbf{0.011} & $-0.007$ (improved) \\
DeepSeek-Chat     & 0.017 & 0.017 & $\approx 0$ (unchanged) \\
Gemini 2.5 Flash  & 0.038 & 0.035 & $-0.003$ (slight improvement) \\
\bottomrule
\end{tabular}
\end{table}

A stability gradient emerges across models. Claude Haiku and DeepSeek-Chat exhibit the lowest variance under P01 ($\text{SD} = 0.018$ and $0.017$ respectively), indicating comparatively reproducible coding behavior. Gemini 2.5 Flash shows the highest instability ($\text{SD} = 0.038$), with individual run $\kappa$ values spanning a range of 0.105 ($0.363$--$0.468$). Multi-shot prompting reduces variance for Claude Haiku ($\text{SD}: 0.018 \to 0.011$), while DeepSeek-Chat and Gemini exhibit marginal changes.

These findings reinforce that reporting a single-run agreement score is methodologically insufficient. Repeated executions should be considered a minimum standard when evaluating LLM-based qualitative coding systems. \revised{Although the SD of $\kappa$ is modest for Claude~Haiku and DeepSeek-Chat ($\approx 0.017$), Gemini~2.5~Flash's individual run $\kappa$ spans a range of 0.105 ($0.363$--$0.468$)---a difference that crosses the Fair--Moderate boundary and would materially affect interpretation if only a single run were reported. Furthermore, ten independent runs constitute the minimum sample size required for a valid Wilcoxon Signed-Rank Test ($n \geq 10$), which underpins the inferential analysis in RQ2.}

\subsection{RQ4 -- Bias in Psychological Safety Coding}

Figure~\ref{fig:ecbias} and Table~\ref{tab:ecbias} present bias ratios for the ``Expressing Concerns'' category across models and prompt configurations. Table~\ref{tab:fullbias} provides a full breakdown of bias ratios across all seven categories.

\begin{figure}[t]
  \centering
  \includegraphics[width=\columnwidth]{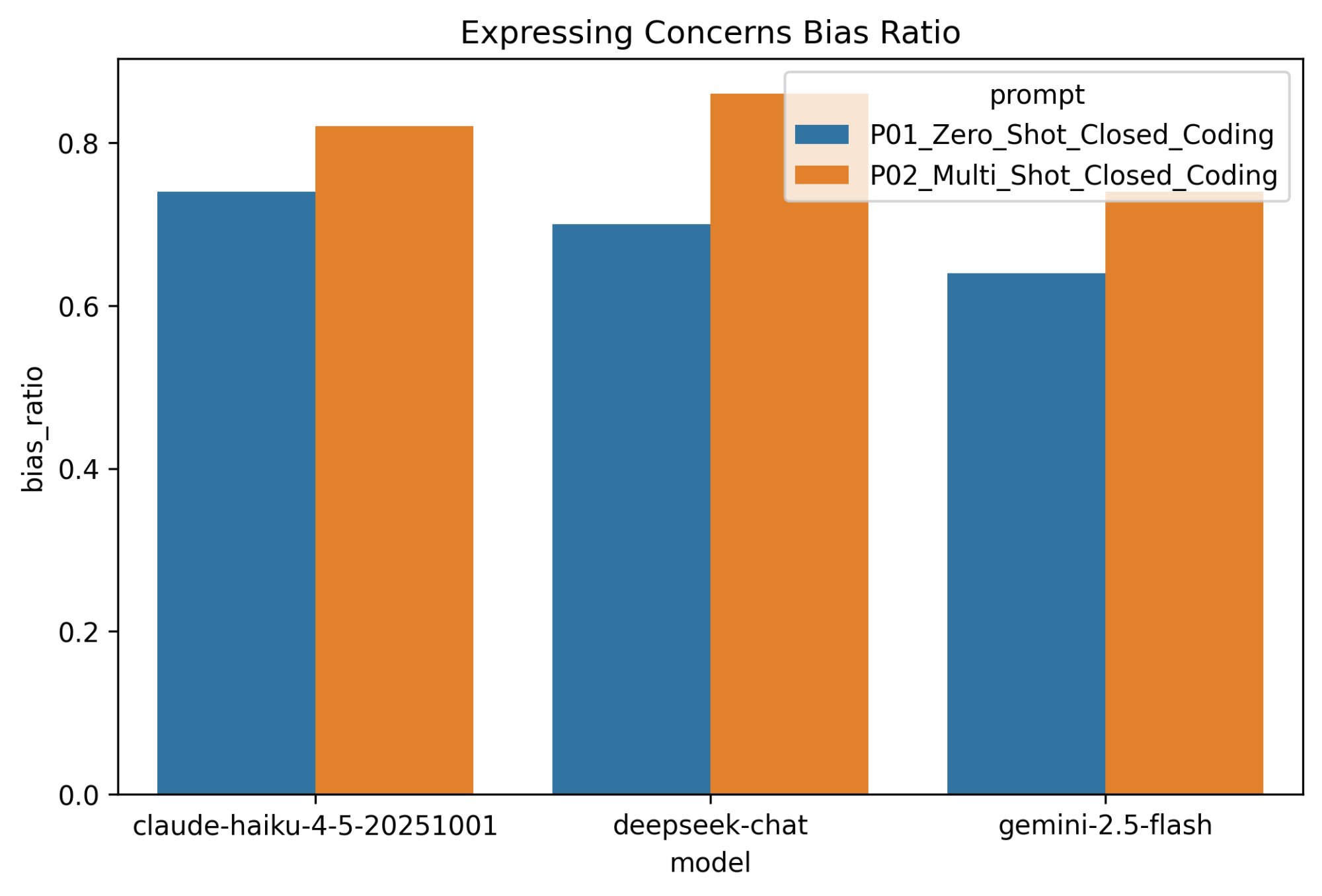}
  \caption{``Expressing Concerns'' prediction bias ratio across models under P01 (zero-shot) and P02 (multi-shot). Values below 1.0 indicate under-prediction relative to the Gold Standard ($n=50$, 43.1\%). All three models under-predict this category.}
  \label{fig:ecbias}
\end{figure}

\begin{table}[t]
\centering
\caption{``Expressing Concerns'' Bias Ratios Across Configurations}
\label{tab:ecbias}
\footnotesize
\setlength{\tabcolsep}{3.5pt}
\begin{tabular}{@{}lrrrr@{}}
\toprule
\textbf{Model} & P01 Count & P01 Ratio & P02 Count & P02 Ratio \\
\midrule
\rowcolor{rowgreen}
Human GS         & 50 & 1.00 & 50 & 1.00 \\
Claude Haiku     & 37 & 0.74 & 41 & 0.82 \\
DeepSeek-Chat    & 35 & 0.70 & 43 & 0.86 \\
Gemini 2.5 Flash & 32 & 0.64 & 37 & 0.74 \\
\bottomrule
\end{tabular}
\end{table}

\begin{table}[t]
\centering
\caption{Full Category Bias Ratios Under P01 (Zero-Shot)}
\label{tab:fullbias}
\footnotesize
\setlength{\tabcolsep}{3pt}
\begin{tabular}{@{}lrrrrrrr@{}}
\toprule
\textbf{Model} & \textbf{EC} & \textbf{DS} & \textbf{RC} & \textbf{SH} & \textbf{DAE} & \textbf{SNF} & \textbf{AM} \\
\midrule
\rowcolor{rowgreen}
Human GS         & 1.00 & 1.00 & 1.00 & 1.00 & 1.00 & 1.00 & 1.00 \\
Claude Haiku     & 0.74 & 0.69 & 1.14 & 1.20 & 2.40 & \textbf{4.75} & 0.67 \\
DeepSeek-Chat    & 0.70 & 0.57 & 1.29 & 2.80 & 1.80 & \textbf{4.75} & 0.33 \\
Gemini 2.5 Flash & 0.64 & 0.74 & 0.93 & 2.20 & 2.40 & \textbf{5.25} & 0.33 \\
\bottomrule
\multicolumn{8}{l}{\textit{EC=Expressing Concerns, DS=Disagreeing/Suggestions, RC=Recommending}}\\
\multicolumn{8}{l}{\textit{Changes, SH=Seeking Help, DAE=Drawing Attention to Errors,}}\\
\multicolumn{8}{l}{\textit{SNF=Sharing Negative Feedback, AM=Admitting Mistakes.}}\\
\end{tabular}
\end{table}

The bias analysis reveals a multi-directional pattern. Most notably, ``Expressing Concerns''---the dominant category (43.1\%)---is \emph{under}-predicted by all three models under P01 (ratios: 0.74, 0.70, and 0.64). Multi-shot prompting partially corrects this, raising ratios to 0.82, 0.86, and 0.74 respectively, but EC remains under-predicted across all configurations.

The most severe over-prediction bias is found in ``Sharing Negative Feedback'' (SNF), with ratios of $4.75\times$ (Claude Haiku and DeepSeek-Chat) and $5.25\times$ (Gemini 2.5 Flash) under P01. ``Drawing Attention to Errors'' (DAE) is similarly over-predicted ($1.80\times$--$2.40\times$). This pattern suggests all three models conflate semantically adjacent categories---EC, SNF, and DAE all involve challenging or critical communicative acts---by over-labeling statements as SNF or DAE rather than EC. Under P02, SNF ratios remain elevated ($4.50\times$, $3.25\times$, $5.25\times$), confirming that this is a model-level semantic prior rather than a prompt-level artifact. ``Admitting Mistakes'' and ``Disagreeing with Suggestions'' are consistently under-predicted across all configurations.

\subsection{Summary of Findings}

Across all experiments, four key results emerge. First, all three models achieve fair to moderate agreement with human qualitative coding ($\kappa = 0.33$--$0.44$), confirming partial but limited reproducibility. Second, multi-shot prompting significantly improves agreement for Claude Haiku ($p = 0.004$) and shows a positive trend for Gemini 2.5 Flash, while leaving DeepSeek-Chat unaffected. Third, Gemini 2.5 Flash exhibits the highest run-to-run variance, reinforcing the need for repeated-run evaluation. Fourth, systematic bias is present in all models: ``Sharing Negative Feedback'' is severely over-predicted (up to $5.25\times$), while ``Expressing Concerns'' is consistently under-predicted---a pattern that persists across both prompt configurations.

\section{Discussion}

\textbf{RQ1 -- LLMs Achieve Partial but Limited Agreement.}
The fair to moderate $\kappa$ scores ($0.33$--$0.44$) confirm that LLMs can partially reproduce human qualitative coding of psychological safety statements, but a substantial proportion of disagreements persist. Per-class F1 analysis reveals that performance is not uniform: models perform best on frequent and semantically distinct categories (``Disagreeing with Suggestions or Ideas'') and worst on rare or overlapping ones (``Sharing Negative Feedback'', ``Admitting Mistakes''). These findings align with prior work showing that LLMs can approximate human annotation in well-defined tasks~\cite{bano2024}, while confirming their reliability is insufficient for fully autonomous qualitative analysis.

\textbf{RQ2 -- Prompt Design Differentially Affects Models.}
The significant improvement for Claude Haiku ($p = 0.004$, $d = 2.41$) suggests that some models benefit considerably from in-context examples, while others---like DeepSeek-Chat---are unaffected. This challenges the assumption that multi-shot prompting offers uniform benefits~\cite{liu2023}. The practical implication is that prompt strategy should be calibrated per model rather than applied uniformly, and researchers should empirically evaluate prompt sensitivity before adopting a fixed strategy.

\textbf{RQ3 -- Stability Warrants Multi-Run Reporting Standards.}
Gemini 2.5 Flash's higher variance ($\text{SD} = 0.038$) versus Claude Haiku and DeepSeek-Chat ($\text{SD} \approx 0.017$) highlights a key methodological concern: a single-run evaluation may yield a misleading picture of reliability. The stability improvement for Claude Haiku under multi-shot prompting ($\text{SD}: 0.018 \to 0.011$) demonstrates that prompt design affects not only mean agreement but also consistency. These findings strengthen the case for standardized multi-run evaluation protocols in LLM-based qualitative coding research.

\textbf{RQ4 -- Bias Is Pervasive and Category-Specific.}
The systematic over-prediction of SNF and DAE alongside under-prediction of EC suggests a shared semantic tendency across models to interpret ambiguous interpersonal statements as expressions of criticism. The EC under-prediction pattern---where models distribute EC-appropriate quotes across SNF and DAE---may reflect semantic overlap among the three categories, all of which involve challenging or negative communicative acts. That these biases persist across both prompt configurations confirms they reflect model-level priors rather than prompt-level artifacts. Bias monitoring across all categories---not only the dominant one---should therefore be a standard diagnostic step in LLM-assisted qualitative coding. \revised{It bears noting that the Bias Ratio measures distributional volume rather than per-instance accuracy; the category-level F1 scores reported in Section~\ref{sec:results} provide the complementary instance-level perspective.}

\textbf{Actionable Guidelines.}
Four actionable guidelines emerge from this study: (1)~prompt strategy should be calibrated per model; multi-shot prompting offers significant benefits for some models but not others; (2)~multi-run evaluation should be adopted as a minimum standard; (3)~full category-level bias monitoring should be embedded in any LLM coding workflow with imbalanced label distributions; (4)~minority category performance requires special attention, as macro-averaged metrics may obscure near-zero F1 on rare but theoretically important categories.

\section{Threats to Validity}

\textbf{Construct:} Seven behavioral codes may not capture all nuances of psychological safety. All categories derive from Edmondson's validated framework~\cite{edmondson1999}. Agreement metrics (Cohen's $\kappa$, macro-F1) may not fully capture the nuanced reasoning involved in qualitative interpretation. \revised{Semantic overlap among EC, SNF, and DAE---all involving critical or negative communicative acts---may independently contribute to the observed bias patterns, irrespective of model limitations. Disentangling model-level from scheme-level sources of classification error remains an open challenge and is reflected in the category-level discriminative guidance embedded in the prompts.}

\textbf{Internal:} Prompt wording is a primary validity threat, mitigated by a fixed schema across all configurations. The P01 single-example-per-code design may introduce implicit labeling bias; the specific example selected per category could influence model priors.

\textbf{External:} 116 quotes from two Stack Exchange communities may not represent industrial, organizational, or cross-cultural settings. The generalizability of the observed bias patterns to other qualitative coding frameworks with different label distributions remains an open empirical question.

\textbf{Reliability:} Ten independent runs explicitly address non-determinism; however, ten runs may still underestimate variance for highly stochastic configurations. The stability results should be interpreted as approximate estimates of run-to-run variability rather than precise variance characterizations. \revised{Although \texttt{temperature~=~0} is near-deterministic in practice, it does not guarantee fully identical outputs across runs; the 10-run protocol was adopted precisely to capture residual non-determinism.}

\textbf{Statistical power:} Wilcoxon tests on ten-pair comparisons have moderate power. The significant result for Claude Haiku is robust; the near-significant result for Gemini ($p = 0.084$) should be interpreted with caution and warrants replication with larger run counts.

\revised{\textbf{Gold Standard reliability:} The Gold Standard from Santana et al.~\cite{santana2023} was produced through a multi-researcher consensus process (4.67\% disagreement rate reported); however, no formal inter-rater reliability statistic (e.g., Cohen's~$\kappa$) is reported in the original study. This limits our ability to contextualise the LLM--human agreement scores against a human--human baseline.}

\section{Conclusion}

This study presented a controlled empirical evaluation of three LLMs across two prompt engineering configurations applied to a human-coded Gold Standard of 116 psychological safety quotes from Stack Exchange SE communities. Running each model--prompt pair ten independent times provides empirical grounding for prompt engineering guidelines in LLM-assisted qualitative coding for SE. Key findings: \textbf{(1)}~all three models achieve fair to moderate agreement with human coding ($\kappa = 0.33$--$0.44$); \textbf{(2)}~multi-shot prompting significantly improves agreement for Claude Haiku ($\Delta\kappa = +0.034$, $p = 0.004$), while having no meaningful effect on DeepSeek-Chat; \textbf{(3)}~Gemini 2.5 Flash exhibits the highest run-to-run variance, highlighting the need for multi-run evaluation protocols; \textbf{(4)}~systematic over-prediction of ``Sharing Negative Feedback'' (up to $5.25\times$) and under-prediction of ``Expressing Concerns'' persist across all models and configurations, reflecting model-level semantic priors. These findings motivate investigation of chain-of-thought and contrastive prompting strategies, standardized multi-run reporting requirements, and extension to larger and more diverse datasets. \revised{Zero-shot and multi-shot configurations were selected as they represent the minimal and enriched baselines most directly comparable in the closed-coding literature; chain-of-thought and contrastive prompting are explicitly deferred to future work to maintain a focused experimental comparison. An analysis of inter-LLM agreement---examining whether models converge on the same labels independently of the gold standard---also represents a valuable complementary direction for future investigation.}

\section*{Availability of Artifacts}

The replication package associated with this study is publicly archived on
Zenodo and includes the reference datasets, prompt specifications,
analysis pipeline, and execution scripts used in the experiments.

\begin{quote}
Moaath Alshaikh. (2026). \textit{PROMPT-SE 2026 Replication Package
(v1.0.0)}. Zenodo.
\url{https://doi.org/10.5281/zenodo.20032899}
\end{quote}

The corresponding GitHub repository is available at:
\url{https://github.com/moaathalshaikh/PROMPT-SE-2026-Replication-Package}

\balance
\clearpage
\bibliographystyle{ACM-Reference-Format}
\bibliography{sample-base}

\clearpage
\appendix

\section{Full Prompt Specifications}
\label{app:prompt}
\subsection{P01: Prompt 01 (Zero-Shot)}

\begin{footnotesize}
\begin{verbatim}
You are a software engineering researcher specializing in qualitative
analysis of psychological safety in software development teams.

Your task is to perform qualitative coding of the quotes provided below
by identifying and classifying psychological safety challenges described
in each quote.

The goal is to identify challenges related to psychological safety, such
as interpersonal risk-taking, learning anxiety, defensive behaviors, or
resistance to change.

=== CONTEXT FOR CLASSIFICATION ===

Psychological safety refers to a person's perception that they can take
interpersonal risks in a team environment without fear of negative
consequences such as embarrassment, rejection, or punishment.

=== UNIT OF ANALYSIS ===

Each quote represents a single unit of analysis.

Even if multiple psychological challenges appear in the same quote,
identify and classify only the ONE primary challenge that best represents
the main issue expressed in the quote.

=== INSTRUCTIONS ===

Step 1 - Read and Analyze the Quote
Carefully read each quote and analyze the psychological challenge described.

Step 2 - Classify the Challenge
Assign the quote to ONE of the following categories.

=== CATEGORIES ===

  Expressing concerns
  Recommending changes
  Sharing negative feedback
  Disagreeing with suggestions or ideas
  Drawing attention to errors
  Seeking help
  Admitting mistakes

=== CATEGORY SELECTION RULE ===

Each quote must be assigned to ONE primary category only.

If multiple categories seem applicable, choose the category that best
represents the central psychological challenge described in the quote.

The category value must match exactly one of the seven category names
listed above.

=== AMBIGUITY HANDLING RULE ===

If a quote does not clearly match any category, select the closest
category based only on explicit evidence in the quote.

If no explicit signal is present, choose the category supported by the
strongest textual evidence.

Avoid speculative interpretation of meaning, intent, or context not
stated in the quote.

=== DECISION PRIORITY RULE ===

If a quote explicitly mentions a mistake, bug, defect, or failure,
prioritize the category "Drawing attention to errors" over "Expressing
concerns" or "Sharing negative feedback".

=== EVIDENCE RULE ===

The identified challenge must be grounded strictly in the content of
the quote. Do not infer motivations, intentions, or contextual
information that are not explicitly stated in the quote.

=== CHALLENGE DESCRIPTION RULE === 

The field  ``challenge\_identified `` must contain a concise description
between 5 and 15 words summarizing the psychological challenge.

Prefer neutral, descriptive wording rather than interpretive language.

=== COVERAGE RULE === 

Analyze every quote exactly once.

=== OUTPUT FORMAT (STRICT) ===

Return ONLY valid JSON.

The output must be a JSON array where each element contains the
following fields:
  * id\_quote
  * category
  * challenge\_identified
  * related\_quote

Return the results ONLY as a JSON array.
\end{verbatim}

\subsection{P02: Prompt 02 (Multi-Shot)}

\begin{verbatim}
You are a software engineering researcher specializing in qualitative
analysis of psychological safety in software development teams.

Your task is to perform qualitative coding of the quotes provided below
by identifying and classifying psychological safety challenges described
in each quote.

The goal is to identify challenges related to psychological safety, such
as interpersonal risk-taking, learning anxiety, defensive behaviors, or
resistance to change.

=== CONTEXT FOR CLASSIFICATION ===

Psychological safety refers to a person's perception that they can take
interpersonal risks in a team environment without fear of negative
consequences such as embarrassment, rejection, or punishment.

=== UNIT OF ANALYSIS ===

Each quote represents a single unit of analysis.

Even if multiple psychological challenges appear in the same quote,
identify and classify only the ONE primary challenge that best represents
the main issue expressed in the quote.

=== INSTRUCTIONS ===

Step 1 - Read and Analyze the Quote
Carefully read each quote and analyze the psychological challenge described.

Step 2 - Classify the Challenge
Assign the quote to ONE of the following categories.

=== CATEGORIES ===

  Expressing concerns
  Recommending changes
  Sharing negative feedback
  Disagreeing with suggestions or ideas
  Drawing attention to errors
  Seeking help
  Admitting mistakes

=== ANNOTATED EXAMPLES ===

Example 01
  id\_quote: Quote12
  category: Disagreeing with suggestions or ideas
  challenge\_identified: Team members hesitate to raise concerns about
    effort estimation in Scrum

Example 02
  id\_quote: Quote16
  category: Drawing attention to errors
  challenge\_identified: Recurring performance issues in Front-End team
    despite extra effort

Example 03
  id\_quote: Quote01
  category: Expressing concerns
  challenge\_identified: Difficulties dealing with argumentative and
    uncooperative team members

Example 04
  id\_quote: Quote02
  category: Recommending changes
  challenge\_identified: Encouraging team to adopt agile vertical slicing
    practices

Example 05
  id\_quote: Quote55
  category: Seeking help
  challenge\_identified: Difficulties obtaining support from a reluctant
    experienced engineer

Example 06
  id\_quote: Quote106
  category: Sharing negative feedback
  challenge\_identified: Receiving harsh criticism during a first code review

Example 07
  id\_quote: Quote42
  category: Admitting mistakes
  challenge\_identified: Struggles with delegation and task management as
    a project manager

=== OUTPUT FORMAT (STRICT) ===

Return ONLY valid JSON.

The output must be a JSON array where each element contains the
following fields:
  * id\_quote
  * category
  * challenge\_identified
  * related\_quote

Return the results ONLY as a JSON array.
\end{verbatim}

\end{footnotesize}
\end{document}